\documentstyle[11pt,agums,epsf,graphics]{article}

\def\IR{{\hbox{{\rm I}\kern-.2em\hbox{\rm R}}}}
\def\IH{{\hbox{{\rm I}\kern-.2em\hbox{\rm H}}}}
\def\IC{{\ \hbox{{\rm I}\kern-.6em\hbox{\bf C}}}}
\def\IZ{{\hbox{{\rm Z}\kern-.4em\hbox{\rm Z}}}}

\newcommand{\beq}{\begin{equation}}
\newcommand{\be}{\begin{equation}}
\newcommand{\eeq}{\end{equation}}
\newcommand{\ee}{\end{equation}}
\newcommand{\bea}{ \begin{eqnarray}}
\newcommand{\eea}{\end{eqnarray}\par\par \noindent }
\newcommand{\bean}{\begin{eqnarray*}}
\newcommand{\eean}{\end{eqnarray*}}
\newcommand{\ba}{\beq\begin{array}{lll} }
\newcommand{\ea}{\end{array}\eeq}

\def\IC{ {\rm l\hspace{-1.2ex}C} }    
\def\IZ{{\hbox{{\rm Z}\kern-.4em\hbox{\rm Z}}}}
\def\IR{{\hbox{{\rm I}\kern-.2em\hbox{\rm R}}}}

\newcommand{\titulo}{Analysis of Water Vapor spatio-temporal structure over the Madrid Area using GPS data}
 
\lefthead{RUFFINI ET AL.}

\righthead{ANALYSIS OF WV SPATIO-TEMPORAL STRUCTURE}

\received{July~30,~1998}
\revised{November~5,~1998}
\accepted{November~12,~1998}

\paperid{}

\authoraddr{ G. Ruffini,  Institut d'Estudis Espacials de Catalunya (IEEC), CSIC Research Unit, Edif. Nexus-204, {Gran Capit\`{a}} 2-4, 08034 Barcelona, Spain. (e-mail:
ruffini@ieec.fcr.es)}

\begin{document}

%
%

\title{\titulo}


%
%

\author{G. Ruffini\altaffilmark{1},  A. Rius\altaffilmark{1},  L. Cucurull\altaffilmark{1}, A. Flores\altaffilmark{1} }
\altaffiltext{1}{Institut d'Estudis Espacials de Catalunya, CSIC Research Unit, E-08034-Barcelona,
Spain.}

%
%

\begin{abstract}
We have analyzed Zenith Wet Delay (ZWD) time series from an experiment over the Madrid (Spain) area obtained
from 5 GPS receivers using two different techniques. In the first case a delay correlation analysis of the ZWD time-series has been carried out. We show that for this small network (with a spatial scale of less than 100 km) the correlation between the time series is very strong, and that using windowing techniques a reliable correlation delay time series can be produced for each pair of sites (10 such pairs are available).  We use this delay time series together with a frozen flow model to estimate the velocity of a passing front, and compare the results to meteorological data and Numerical Weather Prediction output, showing good agreement.  In the second approach, the data is analyzed using Empirical Orthogonal Functions. We demonstrate that the temporally demeaned and normalized  analysis yields information about the passing of fronts, while the spatially demeaned data yields orographic information.  A common second mode highlights the underlying wave behavior. 
\end{abstract}

\begin{article}

\section{Introduction}

If an important goal for the GPS research community has been  to test the limits of the
geophysical measurement techniques  derived  GPS  technology, a now pressing task is to use the newly available data for meteorological studies.  
We will focus here on trying to extract relevant information from the new type of data generated by the GPS  measuring technique.   In a previous publication we discussed the analysis of Zenith Wet Delay (ZWD) and  gradients measured with GPS and Water Vapor Radiometers (WVR) \cite{Ruffini99}. We will now analyze the spatio-temporal structure of the obtained GPS Zenith Wet Delay Time series.  Similar studies have been carried out by \cite{Davis} (although the scale of the network involved, the Swedish permanent GPS network,  is  significantly different, where it was already pointed out that GPS water vapor estimates can be very useful for studying the spatial progress of air masses.   

The  refractivity of the neutral atmosphere at radio frequencies is given approximately
by 
$
 {N}\approx 77.6{P/ T}+3.73\times 10^5 {P_w /  T^2} \equiv  {N^{dry}}+ {N^{wet}} 
$,  where $P$ is the total pressure, $P_w$ is the water vapor partial pressure (both in mb), and $T$ is the temperature (in K). The  equivalent excess path length  corresponding to a ray  crossing the neutral atmosphere is given by \cite{GPS0}
$
\Delta L=10^{-6}\int {N} \, dl + S - {\cal G} 
$. 
Here 
$\cal G$ is the straight-line distance between satellite and receiver,
and $S$ is the geometric path length along the ray. 
In order to estimate Zenith Total Delay  (ZTD)  mapping functions are used, because GPS measurements are not, in general,  in the zenith direction. 
Mapping functions model the dependence of the tropospheric delay   on  satellite elevation, making some assumptions about the tropospheric gas distribution. In the past, azimuthally symmetric models were  employed, and the elevation dependent slant delay  approximated  by ${\Delta L(e)}\approx {\Delta L_z^{dry} m_{dry}(e)}+{\Delta L_z^{wet} m_{wet}(e)}$, where $m_{wet}(e)$ and $m_{dry}(e)$ are  elevation  mapping functions.  Tropospheric delay gradient estimation  is  now possible and routinely carried out \cite{Bar-Sever,MacMillan}.

Two types of tropospheric gradients were considered in \cite{pedro2,Ruffini99}.
Let the atmospheric refractivity be given by
$
N(\vec{\rho},z)\approx N_0(z) +\left. \nabla_{{\rho} } N(z, \vec{\rho}) \right|_{\vec{\rho}=0 } \cdot \vec{\rho}
$, 
where $z$ is the height coordinate and $\vec{\rho}$ the horizontal displacement vector, 
and define $ \left. \nabla_{{\rho} } N(z, \vec{\rho}) \right|_{\vec{\rho}=0 } \equiv \vec{g}( z)$, the horizontal {\em refractivity gradient}. 
On one hand, there is a gradient associated with azimuthal dependencies of delay observations at a GPS receiver.  
This GPS tropospheric slant-delay gradient ({\em sd-gradient}), $\vec{G} =(G_N,G_E) $,  is defined by  the non-azimuthally symmetric delay part in the GPS signal by
$
\Delta D(e,\hat{\rho}) = m_\Delta (e) \cot (e) \, \vec{G} \cdot \hat\rho, 
$
where $\hat\rho= (\cos \phi ,\sin \phi)= \vec{\rho}/ ||{\vec{\rho}}||$ is the azimuth unit vector \cite{Bar-Sever,MacMillan}.  On the other hand,    $\vec{Z}_G$,  the horizontal gradient estimated with the  zenith delays from a network ({\em zd-gradient}) is more closely related to $\vec {g}$.  As discussed in \cite{pedro2}, one can show 
$
\vec{G} = 10^{-6}\int_0^\infty dz\, z\, \vec{g}(z), \:\: \:\:    \vec{Z} =  10^{-6}\int_0^\infty dz\, \vec{g}(z)$.   

  
Here, we   analyze the data from a GPS campaign carried out in the Madrid area during December 1996 in the light of a frozen flow model. It is known from analysis of meteorological data that two humid, cold fronts crossed the network  during the campaign \cite{cucurull}. See also Figure \ref{96des12_small2.ps} below. 

 The basic idea in the frozen flow  model is that the wet refractivity field  propagates like a wave in the presence of a passing front:
\beq\label{slab}
N(\zeta,\vec{\rho},t)=N_0(\vec{\kappa}\cdot\vec{\rho}-\omega t)\cdot e^{-\zeta/h},
\eeq
where the vectors are (2D) surface vectors. 
Here $\zeta$ is a generalized vertical coordinate that depends on the distance from the geoid ($z$) but which  may also depend on the orography, $\zeta=z+\eta(\vec{\rho})$,  and  $h$ is a vertical scale. The velocity associated with this wave is given by $\vec{v}= \hat{\kappa} \omega/ \kappa$. Notice that temporal and spatial gradients are closely related in  this model: 
$
\vec{v}\cdot\nabla_{{\rho}} N= \nabla_{t}N - N  \nabla_{{\rho} } \eta /h $ or, equivalently, $\vec{v}\cdot\nabla W= {d\over dt}W - W  \nabla_{{\rho} } \eta /h $, if the ZWD (denoted by $W$)  is measured at constant   $z$---not the present case.  

Notice also that the refractivity gradient  is closely related to $\vec{\kappa}$ if we assume a geopotentially stratified atmosphere, i.e., if $z=\zeta$, since by equation~\ref{slab}, we have would have   $ \vec{g}=\left. \nabla_{{\rho} } N(z, \vec{\rho}) \right|_{\vec{\rho}=0 }=  \vec{\kappa} N_0'(-\omega t) ) e^{-\zeta/h}- N(-\omega t) ) \nabla_{{\rho} } \eta /h  $, and the last term would drop out.  Unfortunately, this approximation was not correct for our network, where orography plated an important role.
\section{GPS Campaign}
 In the GPS  campaign (December 1996), we deployed 5 Trimble geodetic GPS receiver systems (called ROBL, ESCO, VILA, IGNE and 
VALD) near the Madrid area, Spain, on December 2-15, 1996, with inter-site separations from 
5 to 50 km \cite{pedro} (see Table~\ref{tablepos}).
 GPS observations consisted of data streams of undifferenced 
dual-frequency carrier-phase and pseudo-range measurements  obtained every 
30 seconds.    
 The GIPSY/OASIS-II (v.4)  software 
package \cite{Webb} (Gipsy)  has been used with a Point Positioning strategy to estimate 
  ZTD at the five GPS sites  with a  precision of 5 mm. 
Estimates of the satellite clock corrections and orbits were provided by the IGS and JPL,  as well as consistent earth-rotation parameters.  Gipsy uses a kalman filtering technique to model time-dependent observables, such as the ZTD. 
The tropospheric delay was modeled as a  random walk, $ \sigma^2 = d^2 \cdot t$, with a drift rate of $d= $ 0.25 cm$/\sqrt{\mbox{h}}$.    The drift rate for the gradient parameters was $0.03$ cm$/\sqrt{\mbox{h}}$.  We used a cut-off elevation angle of $7^o$ (see \cite{Ruffini99} for more data processing details).
The dry part of the delay can be estimated well (to less than 0.35 cm in delay) if surface pressure is known to within 1.5 mb.  We have used pressure estimates produced by HIRLAM together with ground measurements, since barometric measurements were not available at all sites.  A conservative estimate of the  accuracy of the pressure data is better than  a  1.5 mb \cite{cucurull}.

\section{Correlation analysis}
Given the frozen flow model in equation~\ref{slab}, the time series of ZWD at the different sites can be approximated by 
\beq \label{main}
W_i(t)=f(t+\tau_i)e^{-\zeta_i},
\eeq
where $W_i(t)$ represents the ZWD at the $i$-th site at time $t$, $\tau_i$ represents a delay relative to some chosen site, and $e^{-\zeta_i}$ is the  scaling factor that should depend partly on the height of the site  (without loss of generality, we  refer this height to VILA's). Thus,  the delay $\tau_{ij}$ between two sites will be given by
$
\tau_{ij}=\vec{x}_{ij}\cdot\vec{k} = \vec{x}_{ij}\cdot\hat{v}/v
$, 
where the inverse velocity  vector $ \vec{k} $  is given by $\hat{v}/v$, the quotient of the velocity unit vector and its norm,  $\vec{x}_{ij}=\vec{x}_{i}-\vec{x}_{i}$, and $\tau_{ij}= \tau_{i}-\tau_{j}$.
The delay is calculated by finding where the expression
$
F[\tau_{ij}]= \langle W^t_i(t)  W^t_j(t+\tau_{ij})\rangle_\Delta.
$, 
attains its maximum.  The $t$-superscript indicates that the time series' have been temporally demeaned and normalized to unit standard deviation, $W^t=(W-\langle W \rangle_t )/ \sigma_t$, where $\sigma_t \equiv \sqrt{ \langle (W- \langle W  \rangle_t  )^2}\rangle_t$. 
Cross-correlations are  found using a time window given by  $\Delta$ (twelve hours were used here).  In Figure \ref{ESCO-ROBL12.0000ps} we see an example of the analysis, where the signal is clearly seen.  Note the presence of   intervals where the correlation is high, and the delay is zero, which will be discussed below. 

\begin{figure}[h!]
\epsffile{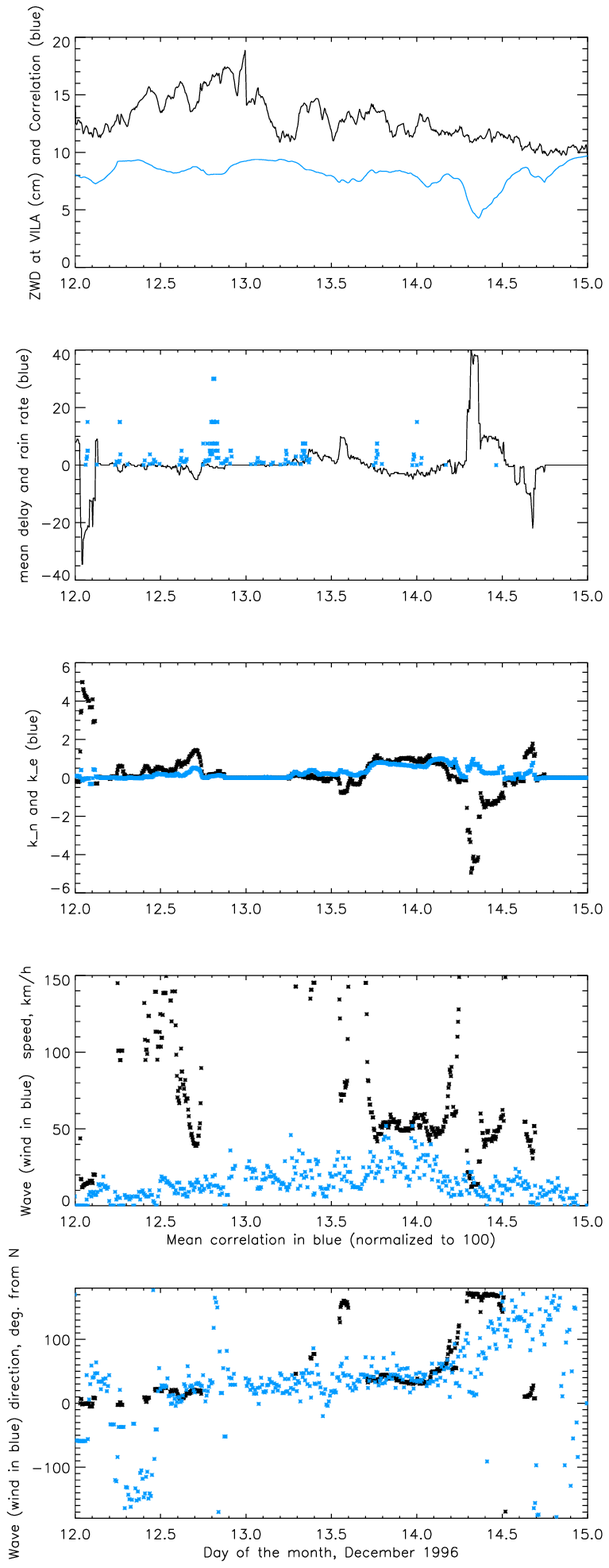} \caption{ \label{winds3.ps} Results of the correlation analysis, and their comparison with meteorological data.  In the top panel we see the ZWD at VILA, and the average correlation function of the analysis (normalized to 10). In the second panel we see the station mean delay from the correlation analysis and the rain rate. In the next panel the estimated $\vec{k}$ is plotted. In the panel below the ground mean speed (in blue) is plotted vs. the estimated wave speed. Finally, the estimated wave direction is plotted against the measured wind direction (in blue) in the bottom panel. }
\end{figure}

The inverse velocity vector $\vec{k}$ has then been estimated, once for each time, by minimizing 
\beq
\chi^2(t)= \sum_{i\neq j} \left( \tau_{ij(t)}- \vec{x}_{ij}\cdot\vec{k}(t)  \right)^2. 
\eeq  
The results are plotted in Figure~\ref{winds3.ps}, and will be discussed below.
In particular, we will try to interpret the existence of intervals where the estimated velocity is infinite, associated with zero delays. 

\section{EOF analysis}
Empirical Orthogonal Function (EOF) decomposition  is a standard tool in multi-variate data analysis. For the task at hand, it will be convenient to think of the ZWD time series at the different sites as a time series of images, $\vec{I}(t)$,  representing at each time the WV content over a 2D network. Given these time series, the goal of EOF analysis is to decompose them  in  orthogonal modes, that is 
\beq
\vec{I}(t)=\sum_{j=1}^5  \lambda_j\vec{I}_j \cdot u^j(t).
\eeq
We have used Singular Value Decomposition, a very useful tool in this context  \cite{Keiner97}. 

Carrying out EOF analysis with the original time series, however, yields a very strong first mode
($\lambda_j =  338.9 ,8.2 ,5.6 ,3.0 , 4.2$) which essentially represents the mean temporal behavior and the exponential vertical behavior.   The two effects, however, are hard to separate and interpret, as spatial and temporal effects tend to get mixed up in the modes. For this reason it is useful to demean and normalize the data first, as we discuss below. 

It is important to emphasize that the most difficult part in EOF analysis is not the numerical  computation of the modes (that is actually fairly simple), but to interpret the results.  For this task we  will make extensive use of the flow model. 
\subsection{Temporal demeaning}
Let us first study the case in which we temporally demean and normalize the time series, since this parallels the correlation approach in the previous section. Recall that we model by equation~\ref{main}. It is the straightforward  to show that
\beq \label{ap1}
W_i^t(t)=  {f(t+\tau_i)-\langle f(t) \rangle_t  \over  s_t}  
\approx 
{ f(t)-\langle f(t) \rangle_t \over  s_t} +{\dot{f}(t)\tau_i \over s_t},
\eeq
where $s_t^2=\left\langle \left(f(t) -\left\langle f(t)\right\rangle_t \right)^2\right\rangle_t$---assuming $\langle \dot{f}(t)  \rangle_t \approx 0$. 
Intuitively,   EOF analysis in this case should  yield information analogous to that of  correlation analysis. We should obtain a spatially homogeneous first mode, with a strong temporal variation, and a second mode with a spatial structure related to the passing front (again, through the delays involved). This spatial structure is directly related to the zd-gradients defined above, except for orographic corrections.   

Another consequence of equation~\ref{ap1} is that $W_i^t(t)=g(t+\tau_i)$. 
This means, for example, in the case of  two GPS sites that the observation matrix will have two columns as follows,
$
A=[g(t) \; g(t+\tau)]
$. 
Carrying out the Singular Value Decomposition yields $ A=UWV^T$ with eigenvalues
$w_\pm=\sqrt{1\pm \langle g(t)g(t+\tau)\rangle_t}$, corresponding $V$ eigenvectors $v_\pm=(1,\pm 1)/\sqrt{2}$, and $U$ eigenvectors $u_\pm=(g(t) \pm g(t+\tau))/w_\pm$.
Thus, we see that the second mode carries  information about the spatial relationships between the delays.  The second spatial eigenvector represents a spatial derivative, while the temporal eigenvector is akin to a time derivative. 

The case with three functions is more difficult to analyze. Let $g_i(t)=g(t+\tau_i)$, and $g_{ij}=\langle g_i(t) g_j(t)\rangle_t$. Let also $s=\sqrt{ g_{12}^2 +g_{23}^2+g_{31}^2 \over 3}$, and $l= g_{12}g_{23}g_{31}$.  Then, the eigenvalues are given by
$
\lambda=1+  2s\left(\cos^{-1} \left( l/  s^{3} \right) + n{2\pi/ 3} \right), 
$
with $n=0,1,2$.  For 
example, if $g_i(t)=e^{-(t-\tau_{i})^2/2\Delta^2}/\Delta\sqrt{4\pi}$,
 we find  $\langle g_i(t) g_j(t)\rangle_t= e^{-\tau_{ij}^2/4\Delta^2}$. 

\begin{figure}[h!]
\epsffile{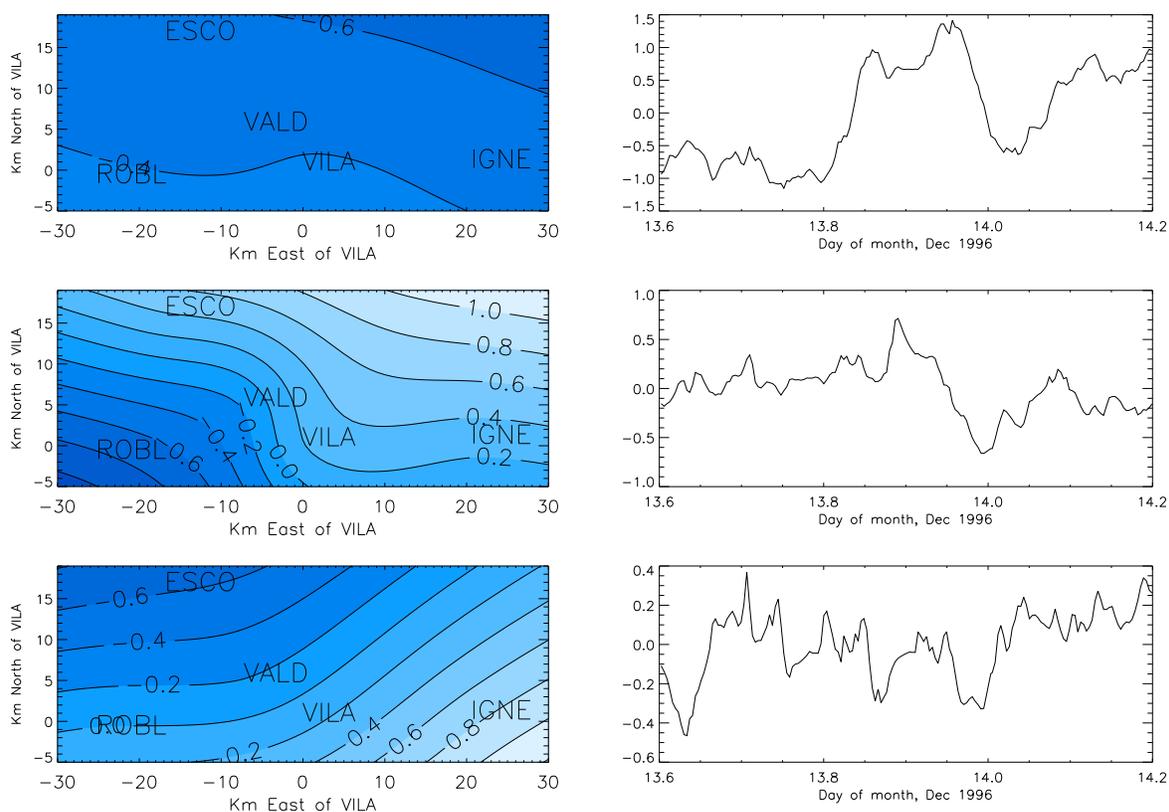} \caption{ \label{temp.eps} The first three modes after temporal demeaning. $\lambda= 10.0 ,      3.3   ,    2.3    ,   1.2   ,    1.4$.  }
\end{figure}

\begin{figure}[h!]
\epsffile{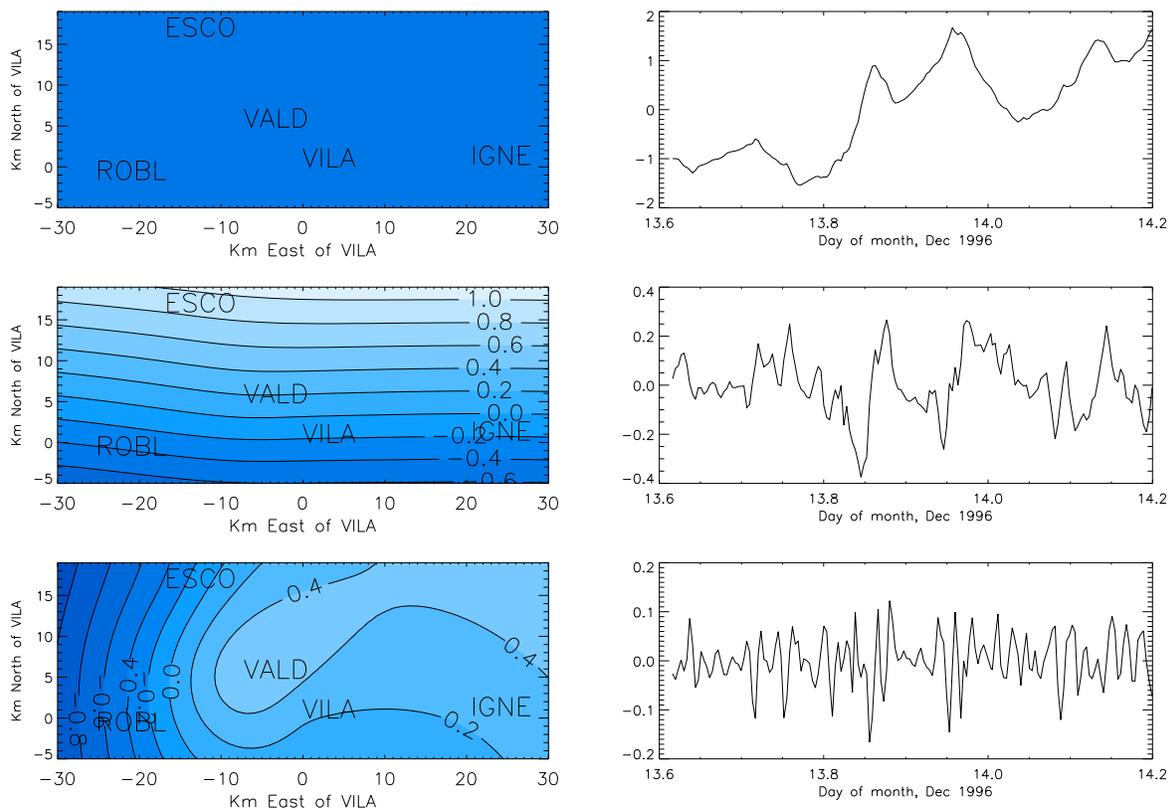} \caption{ \label{temp_sim.eps} The first three modes after temporal demeaning with a simulated 60 km/h north going front. $\lambda=
 12.37 ,   1.6 , 0.6 ,   0.4, 0.0$.  }
\end{figure}

See Figure~\ref{temp.eps} for a representation  of the first three modes using GPS data for one of the periods which involve the passage of a front  (the eigenvalues for the decomposition were $\lambda= 10.0 ,      3.3   ,    2.3    ,   1.4   ,    1.2$).    Our simulations of passing fronts yield precisely this structure. We have generated time series simply by taking one of the real ones and shifting it  in time in a manner conforming to that of a passing front.  An example of the resulting EOF analysis is shown in Figure~\ref{temp_sim.eps}---a north directed front was simulated with a speed of 60 km/h. The eigenvalues in this case are $\lambda=
 12.37 ,   1.6 , 0.6 ,   0.4, 0.0$.   The fact that the last eigenvalue is zero is due to the need to specify only 4 numbers in the simulation (the $\tau_i$'s).
\subsection{Spatial demeaning}
\begin{figure}[h!]
\epsffile{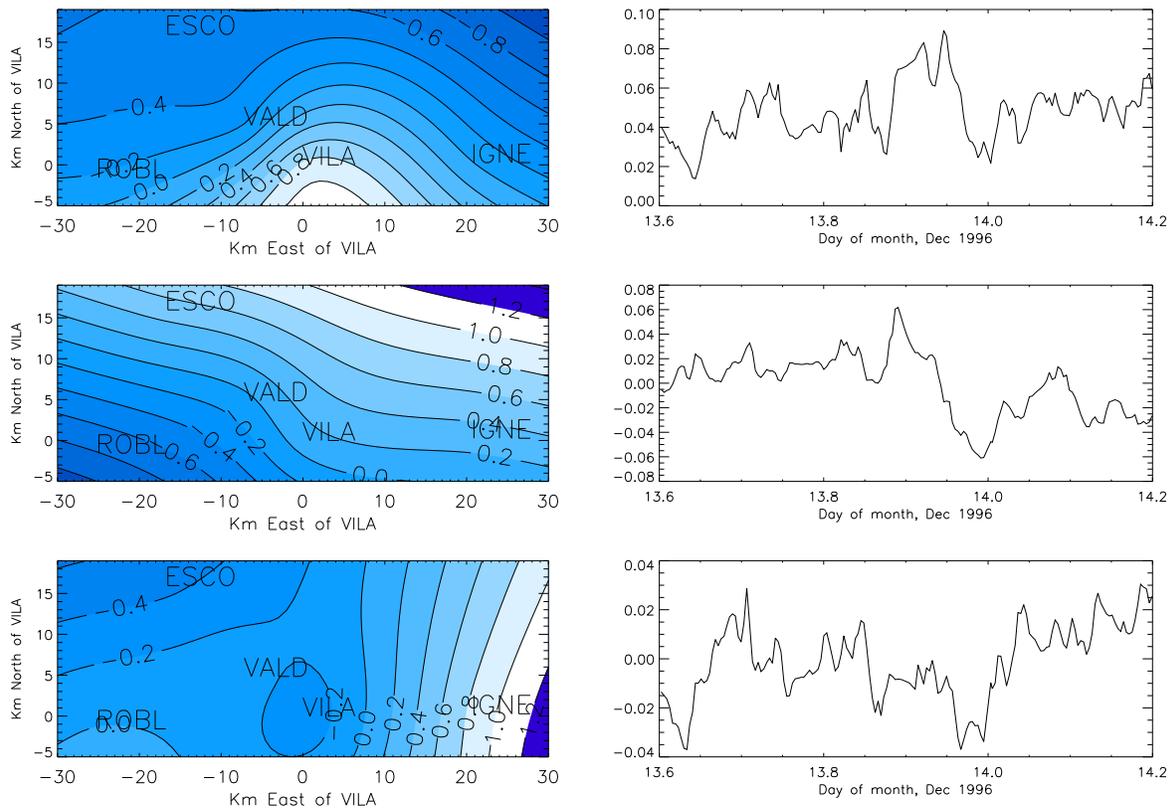} \caption{ \label{spat.eps} The first three modes after spatial demeaning. $\lambda=0.6 ,    0.3  ,   0.2  ,     0.1, 0.0$. } The first mode is closely associated to the orography. 
\end{figure}

In this case, spatial demeaning and normalization to unit variance are carried out prior the EOF decomposition, $W^x=(W-\langle W \rangle_x )/ \sigma_x$, with $\sigma_x=\sqrt{ \langle (W- \langle W \rangle_x  )^2}\rangle_x$.  The $x$ subscript means that statistics are to be computed using the site index $i$. 
From  equation~\ref{main}, we  find,   
\beq \label{ap2}
W^x \approx {e^{-\zeta_i}-\langle e^{-\zeta_i} \rangle_x \over s_x} + {d (\ln f(t))\over dt} \cdot { \tau_i e^{-\zeta_i}-\langle \tau_i e^{-\zeta_i} \rangle_x \over s_x}, 
\eeq
where $s_x=\sqrt{ \langle ( e^{-\zeta_i/h}- \langle e^{-\zeta_i/h}\rangle_x  )^2}\rangle_x$. Hence, 
up to normalization, we may expect modes such as   $v_{0i}\approx e^{-\zeta_i}-\langle e^{-\zeta_i} \rangle$, with very little temporal variation,  and $v_{1i}\approx  \tau_i e^{-\zeta_i}-\langle \tau_i e^{-\zeta_i} \rangle$, with a temporal variation associated again to the time derivative of $f(t)$. The first mode should be a reflection  of  the orography of the network, and the next modes should encode the delay structure associated with the passing front.   See Figure~\ref{spat.eps} for an illustration of the first 3 modes.  The resulting eigenvalues were $\lambda=0.6 ,    0.3  ,   0.2  ,     0.1, 0.0$.  Note here the disappearance of the last mode.



\subsection{Spectral analysis}
The spatially interpolated time-series can also be analyzed spectrally.  This analysis, however, is not simple to carry out or interpret, as we will see. 

Spectral anaysis of the second EOF mode between  the times of 13.6 and 14.2  days reveals peaks at the harmonic spatial frequency with several peaks in temporal frequencies, including one at 0.7 per hour. This  leads to a north-east velocities of around 40 km/h. 
 The measured average surface wind speed was actually 25 km/h, and wind  direction was 42 degrees. 


What should we expect from this analysis?  A simple model for a traveling wave is given by a gaussian wave-packet, $Q=e^{ -(kx-wt)^2}$. The Fourier transform is given by
\beq
F(k,\omega)= {\sqrt{\pi}\over k k'} e^{-k^2/4k'^2} \delta({\omega'/ k'}-{\omega/ k} ).
\eeq
For a  more general wave in a non-dispersive medium, $Q=Q(\vec{k}\cdot \vec{x}-wt)$, we obtain a similar result.
The salient feature is a diagonal spectrum along the constant velocity line ${\vec{k}/ \omega}={\vec{k'}/ \omega'}$.  In 2+1  dimensions the result is 
\beq
F(\vec{k},\omega)=  \delta^2({\vec{k}/ \omega}-{\vec{k'}/ \omega'} ) \int e^{i\omega u} Q(u)du.
\eeq
As may be expected, perfect simulations with the required spatial density illustrate this behavior, yielding a strong diagonal feature. 
Using the simulated data mentioned above, however,   we find a north-direction-time spectrum with an axis of symmetry defined by a  constant velocity vector, but not a  diagonal spectrum.  This is because the simulation, although perfect at the station sites, loses coherence when interpolated in a uniform grid, and the diagonal feature seems to be very unstable. The east direction does conform to the $k_e=0$ equation.  The resulting spectra for the north simulation and the real time series are very similar, however, including the $k_e=0$ equation. 
We conclude that spectral analysis of the raw time series are of limited use.


\section{Conclusions}

We have shown here that cross-correlation and EOF analysis can be  very useful tools for the detection of passing wet/dry fronts in small to medium size networks.  

A striking feature of the EOF analysis is the similarity in the secondary modes in the spatial and temporal demeaning approaches. As has been discussed before (see \cite{Keiner97} and references therein), this is a feature of modes whose  temporal oscillations create spatial gradients, as is the case in the  frozen flow model---a simple wave model. 
 
\begin{figure}[h!]
\epsffile{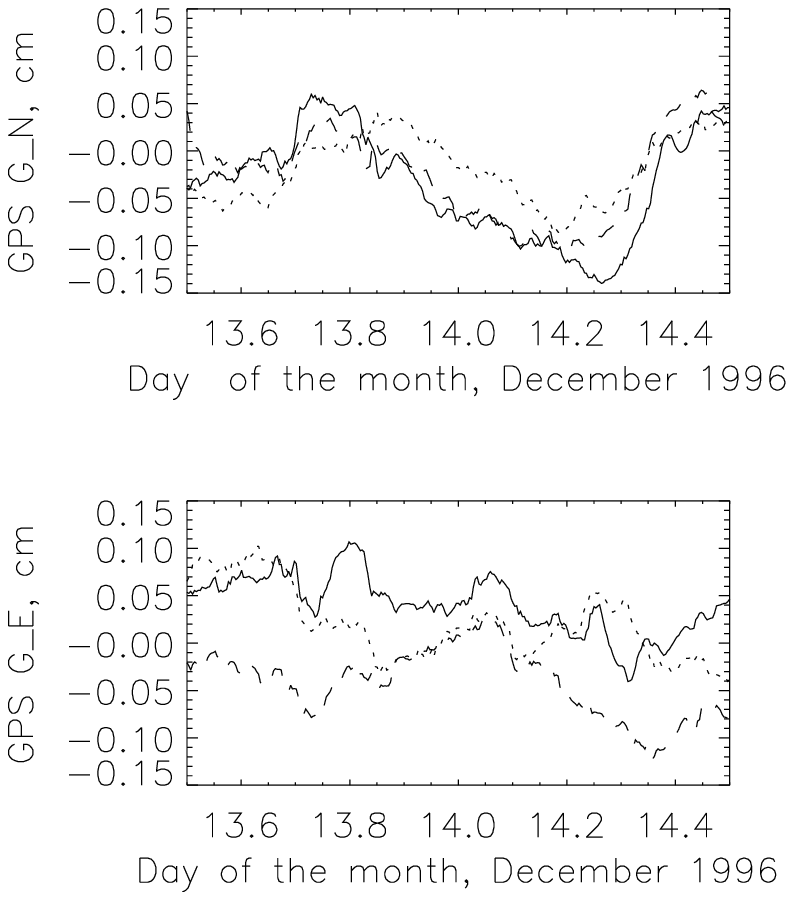} \caption{ \label{grads_gps_new.eps} Gradients obtained with GPS: VILA (solid), ESCO (dots), ROBL (dashes). }
\end{figure}

We can also compare our results with those in \cite{Ruffini99}. We can see there the passage of the front detected here in the form of the obtained sd-gradients (see Figure~\ref{grads_gps_new.eps}), although it is hard at this point to make very quantitative statements. The signal seems to appear in the two analysis, however, as we can see a south pointing gradient. This in, in  effect, a comparison of zd-gradients with sd-gradients. If the exponential law were exact we would see of course a match between the two.  This is not the case, however,  because orography  plays an important role. 

\begin{figure}[h!]
\epsffile{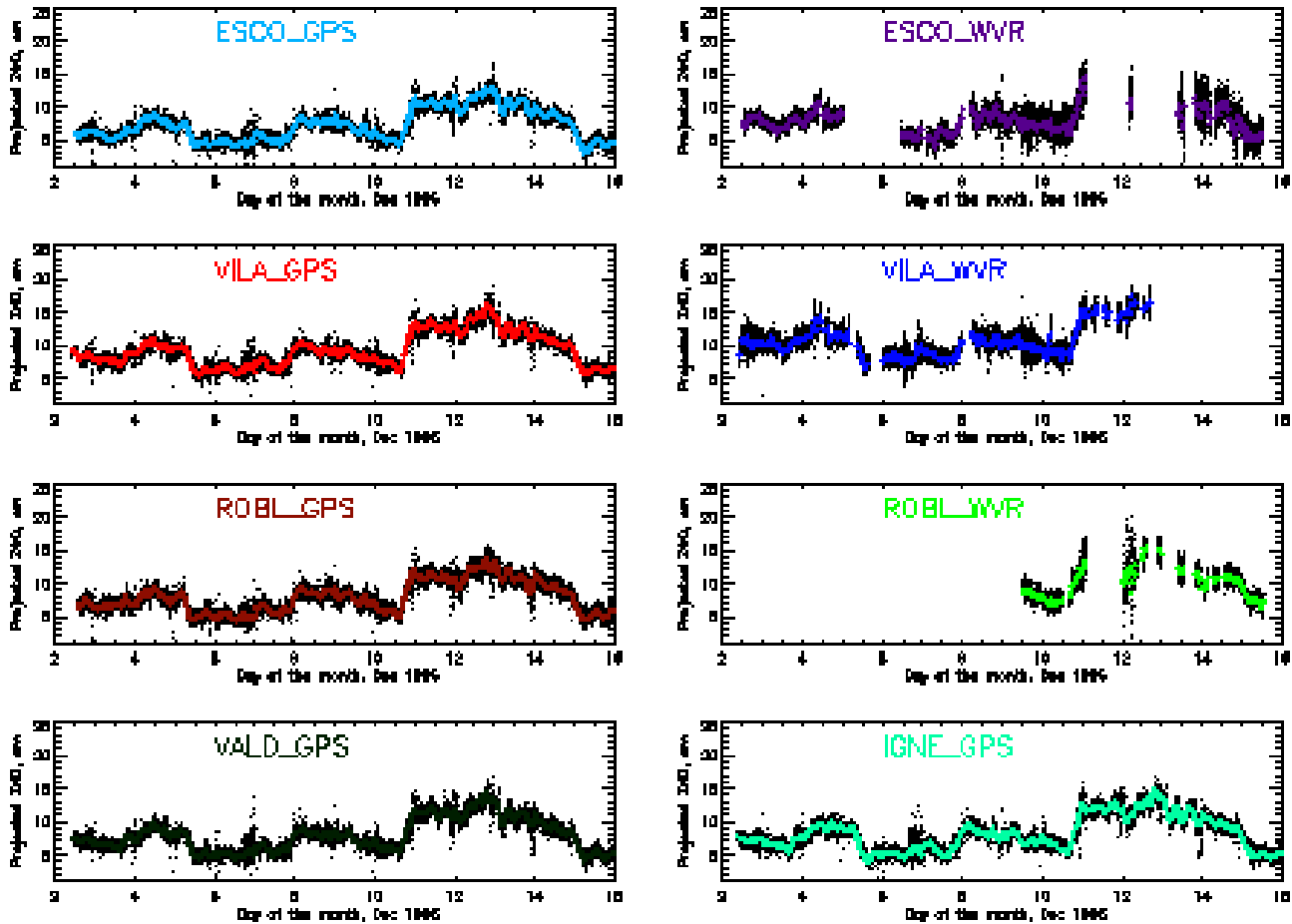} \caption{ \label{All_fitted_madr.ps} The ZWD time series for the GPS and WVR during the campaign. The periods during which no WVR data is available are associated with rain events, as the WVR cannot work in wet conditions.  }
\end{figure}

We hypothesize that  infinite propagation speeds associated  the zero delay correlations  (see Figure~\ref{All_fitted_madr.ps}) are related to rain events. For instance, we can imagine that at some time  it is raining at all sites simultaneously: a  drop in WV will be recorded at all sites, yielding high correlation with zero delays.  Any phenomenon that can change the WV content estimates in the network at the same time will produce this effect, however, so the conclusion is not warranted. We have plotted rain rate measurements in the Figure as well, for comparison. The peak of rain rate does seem to be associated with the zero delay period and a  ZWD drop during that time.

 \acknowledgments 
This work supported by Spanish Climate  CICYT grant CLI95-1781 and by the  WAVEFRONT project,  funded by the European Commission Environment and Climate
Program (EC Contract ENV4-CT96-0301). 
We thank Caltech/JPL for the GIPSY/OASIS--II package.  The Spanish IGN loaned and operated the receivers. GR is grateful to Jordi Vil\`a for useful comments and constructive criticism.
 
   %
   %
   %

%
%

   %
   %
   %

\end{article}

\begin{table}   
{\small
\begin{tabular}{|c|c|c|c|}\hline 
{\bf site} & {\bf $ \Delta X_N$} &   {\bf $ \Delta X_E$} &  {\bf  $\Delta X_Z$ }\\ \hline 
ESCO   &     15.964    &  -16.834& 430  \\
 IGNE  &   0.279    &   20.576& 119\\
 ROBL  &   -1.558   &  -25.256& 181\\
VALD    &   4.852    &  -7.158&197\\
 VILA  &    0.000  &   0.000 & 0\\
\hline 
\end{tabular}
}
%
\caption{\label{tablepos} Positions of the receivers with respect to VILA (all in km except $\Delta X_Z$, in meters.}
\end{table} 

\vfill

\clearpage
\clearpage
\begin{figure}[h!]
\epsffile{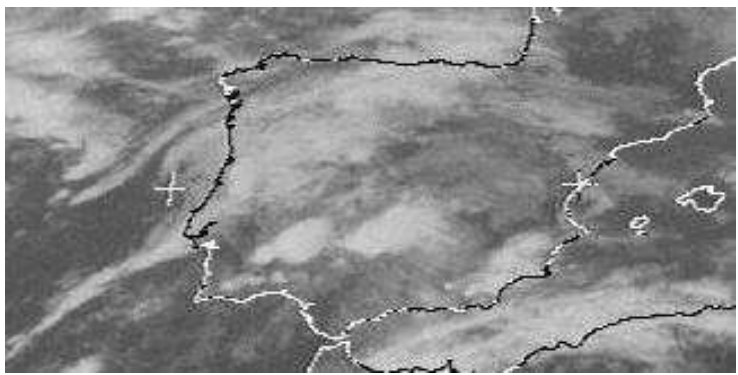} \caption{ \label{96des12_small2.ps} A Meteosat IR photo for December 12 1996 noon UTC (MET5 12 DEC 1996 1200 IR1 D2).  }
\end{figure}

\begin{figure}[h!]
\epsffile{ESCO-ROBL12.0000ps} \caption{ \label{ESCO-ROBL12.0000ps} The delay structure between ESCO and ROBL (correlation is shown dashed and normalized to 100 for graphing purposes). The fact that the delays are negative mean, in the convention used, that ROBL detected the changes before ESCO. In the bottom panel ROBL is shown dashed. }
\end{figure}

\end{document}